\documentclass[conference]{IEEEtran}
\usepackage[utf8]{inputenc}
\usepackage{cite}
\usepackage{graphicx}
\usepackage{amsmath}
\usepackage{mathrsfs,amsthm,amsmath,amssymb}
\usepackage{textcomp}
\usepackage{graphicx,multirow}
\usepackage{balance}
\usepackage{cite}
\usepackage{float}
\usepackage{gensymb}
\usepackage{eso-pic}
\usepackage{algorithm}
\usepackage{algorithmic}
\usepackage[bookmarks=false]{hyperref}
\usepackage{academicons}
\usepackage{xcolor}
\usepackage{ulem}

\makeatletter
\newcommand\subparagraph{%
  \@startsection{subparagraph}{5}
  {\parindent}
  {3.25ex \@plus 1ex \@minus .2ex}
  {-1em}
  {\normalfont\normalsize\bfseries}}
\makeatother
\usepackage{titlesec}
\let\subparagraph\relax

\usepackage{titlesec}
\usepackage{scalerel}
\usepackage{tikz}
\usepackage{subcaption}

\usepackage[english]{babel}

\newcommand{\mb}{\mathbf}

\DeclareMathOperator{\diag}{diag}

\usetikzlibrary{svg.path}

\definecolor{orcidlogocol}{HTML}{A6CE39}
\tikzset{
  orcidlogo/.pic={
    \fill[orcidlogocol] svg{M256,128c0,70.7-57.3,128-128,128C57.3,256,0,198.7,0,128C0,57.3,57.3,0,128,0C198.7,0,256,57.3,256,128z};
    \fill[white] svg{M86.3,186.2H70.9V79.1h15.4v48.4V186.2z}
                 svg{M108.9,79.1h41.6c39.6,0,57,28.3,57,53.6c0,27.5-21.5,53.6-56.8,53.6h-41.8V79.1z M124.3,172.4h24.5c34.9,0,42.9-26.5,42.9-39.7c0-21.5-13.7-39.7-43.7-39.7h-23.7V172.4z}
                 svg{M88.7,56.8c0,5.5-4.5,10.1-10.1,10.1c-5.6,0-10.1-4.6-10.1-10.1c0-5.6,4.5-10.1,10.1-10.1C84.2,46.7,88.7,51.3,88.7,56.8z};
  }
}

\newcommand\orcidicon[1]{\href{https://orcid.org/#1}{\mbox{\scalerel*{
\begin{tikzpicture}[yscale=-1,transform shape]
\pic{orcidlogo};
\end{tikzpicture}
}{|}}}}

\DeclareMathOperator{\re}{Re}

\graphicspath{ {Images/} }
\IEEEoverridecommandlockouts

\begin{document}
\setlength{\parskip}{5pt}
\setlength{\abovedisplayskip}{5pt}
\setlength{\belowdisplayskip}{5pt}


\title{Space MIMO: Direct Unmodified Handheld to Multi-Satellite Communication} 

\author{Yasaman~Omid,  
Zohre~Mashayekh Bakhsh, Farbod~Kayhan, Yi~Ma, Rahim~Tafazolli
 
 \thanks{All authors are with Faculty of Electronics and Physical Sciences, University of Surrey, UK. e-mail: (y.omid, z.mashayekhbakhsh, f.kayhan, y.ma, r.tafazolli)@surrey.ac.uk.}

}

\maketitle
\begin{abstract}
This paper examines the uplink transmission of a single-antenna handsheld user to a cluster of satellites, with a focus on utilizing the inter-satellite links to enable cooperative signal detection. Two cases are studied: one with full CSI and the other with partial CSI between satellites. The two cases are compared in terms of capacity, overhead, and bit error rate. Additionally, the impact of channel estimation error is analyzed in both designs, and robust detection techniques are proposed to handle channel uncertainty up to a certain level. The performance of each case is demonstrated, and a comparison is made with conventional satellite communication schemes where only one satellite can connect to a user. The results of our study reveal that the proposed constellation with a total of $3168$ satellites in orbit can enable a capacity of $800$ Mbits/sec through cooperation of $12$ satellites with and occupied bandwidth of 500 MHz. In contrast, conventional satellite communication approaches with the same system parameters yield a significantly lower capacity of less than $150$ Mbits/sec for the nearest satellite.
\end{abstract}
\begin {IEEEkeywords}
Capacity, LEO Satellite, Mega Constellations,  Multi-Satellite Cooperative Communication.   
\end{IEEEkeywords}
\section{Introduction}
\label{intro}
\IEEEPARstart{S}{atellite}  networks, particularly those comprised of low earth orbit (LEO) satellites, are widely considered as a promising technology for achieving global connectivity. These networks have the potential to connect remote and under-served areas, consequently providing critical access to communication services.
For instance, 
Globalstar and Iridium offer voice and data connectivity to satellite phones and messenger devices but do not provide broadband internet services. Alternatively, OneWeb and SpaceX provide broadband internet access with very small-aperture terminals (VSATs).
A high-throughput connection between satellites and unmodified handheld (UH), such as cellphones, has the potential to enable seamless global connectivity. Despite the benefits of UH-satellite communication, the achievable capacity is limited by the user terminal's (UT) antenna size and gain and the power constraints at both ends.
One potential solution to overcome the limitations of UH antennas is to employ large antennas on satellites, e.g. the AST Bluewalker 3 with its  $64\ \text{m}^2$ antenna. However, this technology is still in the experimental phase and has several drawbacks. These include problems related to space debris, high implementation and launch costs, and probable interference with astronomical observations.
Another possible solution is to enhance data transfer rates through cooperative satellite operations. By doing so, the requirement for larger antennas could be relaxed, and the data throughput within the same constellation could be enhanced.
This technique, which involves collaboration for joint detection/precoding, has only been examined in a limited number of recent publications, such as \cite{abdelsadek2022distributed}, and from now on is referred to as space multi-input multi-output (Space MIMO). To the best of our knowledge, Space MIMO communication has not yet been thoroughly studied in the literature, particularly in the uplink scenario. Thus, focusing on the uplink, in this paper, we study the possibility of a high throughput link for direct UH to a cluster of satellites that cooperatively detect the transmitted signal.

The MIMO technology is a promising technique which enables increased data rates, improved reliability, and better coverage in challenging environments. The use of MIMO technology in satellite communication was reviewed in \cite{arapoglou2010mimo} including both fixed and mobile systems. While several techniques have been proposed to enhance the capacity of MIMO satellite systems, there is currently no research that explores the use of MIMO for direct communication between a mobile user and multiple satellites.

The majority of  the literature on using MIMO in satellite communication focuses on enhancing capacity through other means, rather than considering satellite cooperation. 
In \cite{li2021analysis}, a distributed satellite MIMO setup was proposed to improve the spectral efficiency without increasing the transmit power by using a uniform linear array (ULA) at the ground side. The satellites were positioned far enough apart to allow for a distributed MIMO setup. The study focused only on the downlink scenario for geostationary earth orbit (GEO) satellites. 
The authors in \cite{ramamurthy2016mimo} investigated a comparable system scenario by {analyzing the return link (from UT to anchor station), where each satellite payload is a non regenerative amplify and forward relay. A Rician channel model was examined for MIMO uplink, whereas in downlink two SISO channel was employed, considering only LOS propagation channel model.} 
The authors of \cite{schwarz2019mimo} also investigated a similar system considering the effect of the positions of the ground antenna and the satellite antenna, as well as the perturbations of GEO satellites, in both uplink and downlink scenarios using a flat fading model for MIMO channels.
In \cite{peng2021channel}, the UT movement was examined in a downlink scenario, where each channel coefficient was modelled by a four-state Markov transition process with LOO distribution. This study compared the performance of different transmission schemes, such as dual satellite dual polarization (DSDP), dual satellite single polarization (DSSP), single satellite single polarization (SSSP), and single input single output (SISO). It was revealed that DSDP provides the best bit error rate (BER) performance among all tested schemes.
In \cite{hofmann2017multisatellite}, a mobile MIMO ultra-high frequency (UHF) channel was examined by considering two GEO satellites as transmitters and a uniform circular array (UCA) as a receiver. The MIMO bandwidth efficiency was analysed by rotating the antenna array. The measurement results revealed that a four-element UCA maintains high bandwidth efficiency, regardless of the receiver's spatial orientation, making it the preferable option.
In \cite{dou2014cooperative}, a cooperative beamforming transmission technique between UTs and GEO satellites was proposed. The study derived a ground user selection criterion for user cooperation, and their method resulted in significant improvements in achievable rates for both cooperative and non-cooperative systems. Compared to a single high-power satellite, the cooperative system achieved twice the maximum rate, and the non-cooperative system attained a rate 1.7 times higher.

Some recent studies have explored the advantages of satellite collaboration through inter-satellite links. One such study is \cite{richter2020downlink}, which examined a downlink scenario where single antenna LEO satellites communicate with a land terminal with a UCA antenna. Through synchronization and transmission rate adjustments, the satellites cooperate to improve the data rate. The study demonstrated that the achievable data rate through joint transmission is considerably higher than the achievable rate using a single satellite, even when the latter used twice the transmit power.
In \cite{abdelsadek2022distributed}, a novel architecture for LEO satellite networks utilizing distributed massive MIMO (DM-MIMO) technology was proposed to improve service time for users. The satellite-UT channels were modelled by the Rician distribution, and instantaneous CSI was estimated using pilot-based techniques in each satellite. A cooperative precoding technique was proposed where all satellites in a cluster jointly transmit signals to the UTs to minimize interference. The authors aimed to maximize the UT's aggregate downlink data rate by optimizing power allocation and maximizing the UT's service time. Their proposed approach outperformed single satellite approaches in terms of average service time and spectral efficiency.

Motivated by the above, in this paper we design a space MIMO system for the direct UH to satellite transmission, where the satellites cooperate with each other to detect the transmitted signal. Taking the mobility of the UH into account and following 3GPP standard \cite{3GPPrel15} we consider the land mobile satellite (LMS) channel model in this model.
We investigate the problem of mean square error (MSE) minimization, and we consider satellite cooperation in two cases of full CSI and partial CSI, which are compared with each other in terms of capacity, overhead and BER. 
Our numerical results compare the capacity of these two cases with a conventional satellite communication scheme where only one satellite is connected to the UH. Our study considers a constellation of $3168$ satellites and demonstrates that by incorporating satellite cooperation, we can achieve a remarkable capacity of over $800$ Mbit/sec and a BER of under $0.1$ for uncoded BPSK when only $12$ satellites cooperate on signal processing. 
Notice that conventional communication which lacks satellite collaboration can only achieve a capacity of under $150$ Mbit/sec and a BER of over $0.3$ for the nearest satellite.  

The rest of this paper is organized as follows. The system model and channel model is presented in Section \ref{System Model}. Section \ref{Joint Processing} is dedicated to the joint processing techniques in both cases of full and partial CSI. Section \ref{Numerical Results} presents the simulation results and  section \ref{conclusion} concludes the paper.

\section{System Model}
\label{System Model}
The objective of this study is to examine the uplink transmission of a single-antenna UH user to multiple LEO satellites.  
The satellites under consideration are at different orbits and altitudes, causing the distance between users and each satellite to vary over time. At any given time, we denote by $L_{max}$ the number of  satellites that are within a $30^{\degree}$ elevation angle of the users. Notice that $L_{max}$ depends on the constellation and is a function of time. All satellites have both line-of-sight (LOS) and non line-of-sight (NLOS) links with all users. To enhance system performance and increase capacity, a cluster of satellites is formed by utilizing the inter-satellite links, which jointly detect the received signal from the user. 
We order the satellites depending on their distances from the user, i.e. the nearest satellite to the user is considered as satellite 1.
In this paper we select $L$ satellites from the $L_{max}$ visible satellites that have the least distances from the user to form the cluster at each time $t$. Each user has one omni-directional antenna and transmits the signal towards all satellites simultaneously. The inter-satellite links are assumed to be perfectly known by all satellites, and their effect on the shared information can be eliminate.  The receiver noise at the inter-satellite link is considered negligible. The following subsection presents the channel model of  user-satellites links.

\begin{figure}[t]
    \centering
    \includegraphics[scale=0.32]{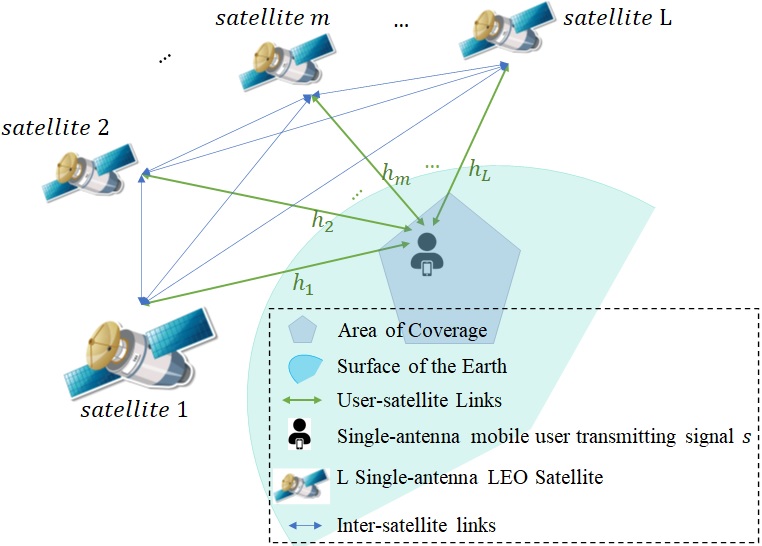}
    \caption{LEO satellite communication system model}
    \label{fig:system model}
\end{figure}

\subsection{Channel Model}
Our flat fading channel employs the two-state model introduced in \cite{3GPPrel15}. This model encompasses both the shadow fading and clutter loss effects. The calculation of the Free Space Path Loss (FSPL) is then modified by including these effects, and can be expressed as ${\text{FSPL}}_m = \frac{4 \pi d_m}{\lambda}$, where $d_m$ represents the distance between the user and satellite $m$, $\lambda$ is the wavelength of the signal which is given by $\lambda = c/f_c$, and $f_c$ denotes the carrier frequency. Doppler shift effect can be compensated at the transmitter \cite{peters2020doppler}.
The two-state model, as defined, categorizes the channel states as either good, with LOS or slightly shadowed conditions, or bad, with severe shadowed conditions. To describe the duration of each state, a semi-Markov model is used, while the LOO distribution characterizes the fading of the channel within each state. This distribution combines both the LOS and non-LOS (NLOS) received signals. 
The channel coefficient between the user and the $m$th satellite is obtained using the following equation:
\begin{align}
h_m = \frac{1}{{\text{FSPL}}_m}(\sqrt{\frac{K}{K+1}}h_{{\text{LOS}}} + \sqrt{\frac{1}{K+1}}h_{\text{NLOS}})
\end{align}
where $K$ is the ratio of the LOS signal power to the mean power of the NLOS signal. The LOS and NLOS channels are denoted as $h_{\text{LOS}}$ and $h_{\text{NLOS}}$ respectively, where $h_{\text{LOS}} = |h_{\text{LOS}}|e^{j\phi_{\text{LOS}}}$ and $h_{\text{NLOS}} = |h_{\text{NLOS}}|e^{j\phi_{\text{NLOS}}}$. The magnitude of $h_{\text{LOS}}$ is log-normally distributed with a mean $M_A$ and standard deviation $\Sigma_A$, and the magnitude of $h_{\text{NLOS}}$ is Rayleigh distributed with an average power $MP$. The phase angles $\phi_{\text{LOS}}$ and $\phi_{\text{NLOS}}$ are uniformly distributed between 0 and $2\pi$.

\section{Joint Processing} \label{Joint Processing}
In this section, we aim to present a joint processing technique among the satellites to minimize the MSE. Here, we assume that each satellite can estimate its own channel, and the CSI information and the received signals are shared among all satellites in the cluster. 
In the absence of inter-satellite link noise, the aggregate capacity of the cluster where each satellite shares its data with all other satellites would be the same as the capacity that would result if all satellites transmitted their data to a single satellite. This is because all satellites would have access to the identical information, and any one of them can perform the detection. Thus, we consider that all satellites share their data with only one satellite which performs the signal processing.
We investigate two cases of full CSI and partial CSI between the satellites. The following subsections elaborate on each of these scenarios.

\subsection{Case 1: full CSI}
In this scenario, we consider that all satellites share their data and instantaneous CSI with only one satellite which performs the signal processing. The overhead of such a system for data sharing is in the order of $\mathcal{O}(L)$ per symbol and for the CSI sharing is in the order of $\mathcal{O}(L)$ per coherence interval.
In more detail, each satellite $m$, $m\in\{1,...,L\}$, shares its estimated channel $\hat{h}_m$, and its received signal $y_m=\sqrt{p}h_ms+n_m$ with a satellite in the cluster which performs the detection. Note that $p=P_TG_TG_R$, where $P_T$ is the transmit power, $G_T$ is the gain of the transmit antenna and $G_R$ is the gain of the receive antenna at the satellite. The additive noise at the receiver is modeled by $n_m\sim\mathcal{CN}(0,\sigma^2)$, where the noise power is calculated by $\sigma^2=k_B\times T\times BW$. Here, $k_B=1.380649\times 10^{-23}$ is the Boltzmann constant, $T=290$ is the noise temperature, and $BW$ is the Bandwidth in Hz. We assume that the channel estimation is imperfect, i.e. $\hat{h}_m=h_m+\Tilde{h}_m$ with $\hat{h}_m$ being the estimation and $\tilde{h}_m$ being the estimation error, where
\begin{align}\label{estimation error distribution}
    &\Tilde{h}_m\sim\mathcal{CN}(0,\sigma_{\tilde{h}_m}^2).
\end{align} 
Now, assuming the joint detection vector is represented by $\mb{v}\in\mathbb{C}^{1\times L}$, the detected signal can be represented by 
\begin{equation}
\hat{s}=\mb{v}\mb{y}=\sqrt{p}s\mb{v}\mb{h}+\mb{v}\mb{n}=\sqrt{p}s\mb{v}(\mb{\hat{h}}-\mb{\tilde{h}})+\mb{v}\mb{n}
\end{equation}
where $\mb{h}=[h_1,...,h_L]^T$, $\mb{\hat{h}}=[\hat{h}_1,...,\hat{h}_L]^T$, $\mb{\tilde{h}}=[\tilde{h}_1,...,\tilde{h}_L]^T$ and $\mb{n}=[n_1,...,n_L]^T$. 

The objective is to find the detection vector that minimizes the MSE. This can be achieved by solving the following optimization problem
\begin{equation}
    \begin{array}{cl}
    \min\limits_{\mb{v}}&  \mathbb{E}\{|\hat{s}-s|^2\},
    \end{array}
\end{equation} 
where the expectation is taken over the additive noise, channel estimation error, and the transmitted symbol. 
The first-order optimality condition can be used to calculate the optimal solution for this optimization problem. Specifically, the gradient of the objective function with respect to the detection vector should be zero at the optimal solution. Therefore, we can set the gradient of the objective function equal to zero and solve the problem for the detection vector. The optimal detection vector can be given by
\begin{equation}\label{5}
    \mb{v}=\sqrt{p}\hat{\mb{h}}^H\left(p \hat{\mb{h}}\hat{\mb{h}}^H+\sigma^2\mb{I}+p\Sigma_h\mb{I} \right)^{-1},
\end{equation}
where $\Sigma_h=diag[\sigma_{\tilde{h}_1}^2,...,\sigma_{\tilde{h}_L}^2]$. 
For further details on (\ref{5}), please refer to appendix \ref{FirstAppendix}.

The capacity in this case is given by the mutual information between $s$ and $\hat{s}$, as
\begin{equation}\label{mutual information}
    I(s;\hat{s})=H(\hat{s})-H(\hat{s}|s),
\end{equation}
where we have $H(\hat{s})\leq \log(\pi e (p\mb{v}\mb{h}\mb{h}^H\mb{v}^H+\sigma^2\mb{v}\mb{v}^H))$,   and $H(\hat{s}|s)=\log(\pi e \sigma^2\mb{v}\mb{v}^H)$. Assuming $s$ has a Gaussian distribution, then $\hat{s}$ would have a Gaussian distribution as well and the capacity for the full CSI scenario is given by 
\begin{equation}\label{rate FC}
    R_{\text{FC}}=\log \left( 1+\frac{p\mb{v}\mb{h}\mb{h}^H\mb{v}^H}{\sigma^2\mb{v}\mb{v}^H} \right).
\end{equation}
The proof is given in Appendix \ref{FirstAppendix}.

\subsection{Case 2: partial CSI}
In this scenario, given that satellite $1$ has the smallest path loss and thus has the strongest channel, we take this satellite as our network controller (NC), and assume that all satellites share their data only with the NC. Note that unlike case 1, the instantaneous estimated CSI of each satellite is not shared with the NC. In this case, a given  satellite  $m$, $m\in\{1,...,L\}$   estimates its own channel imperfectly, and it can perform the following local processing to cancel the effects of channel attenuation
\begin{equation}
    y^{'}_m=\frac{y_m}{\hat{h}_m}=\frac{\sqrt{p}h_ms+n_m}{\hat{h}_m}=\sqrt{p}s-\sqrt{p}s\frac{\tilde{h}_m}{\hat{h}_m}+\frac{n_m}{\hat{h}_m}.
\end{equation}
Then, the satellites $m\in\{2,...,L\}$ send their processed signal $y^{'}_m$ to the NC, along with the following long-term information, i) the average power of the estimated channel $\mathbb{E}\{|\hat{h}_m|^2\}$, ii) the average power of estimation error $\mathbb{E}\{|\tilde{h}_m|^2\}$, and  iii) the average power of inverse estimated channel $\mathbb{E}\{|\frac{1}{\hat{h}_m}|^2\}$. Since these are long-term information, sharing them with the NC does not add to the system overhead. 
Thus, in this case, the system overhead would be in the order of $\mathcal{O}(L)$. After the local processing is performed in each satellite, the processed data is sent to the NC for central processing. The central processing vector is shown by $\mb{w}\in\mathbb{C}^{1\times L}$, and the final detected signal at the NC is presented by $\hat{s}=\mb{w}\mb{y}^{'}$, where $\mb{y}^{'}=[y^{'}_1,...,y^{'}_L]^T$. Similar to case 1, the following optimization problem is solved 
\begin{equation}\label{optimization 2}
    \begin{array}{cl}
    \min\limits_{\mb{w}}&  \mathbb{E}\{|\hat{s}-s|^2\},
    \end{array}
\end{equation}
where the expectation is taken over the additive noise, channel estimation error, the transmitted symbol, and the estimated channels for the satellites $2$ to $L$. 
Given the distribution of the channel estimation error in (\ref{estimation error distribution}), the objective function in (\ref{optimization 2}) can be rewritten as
\begin{align}
    \mathbb{E}\{|\hat{s}-s|^2\}=&p\mb{w}\mb{1}\mb{1}^T\mb{w}^H-\sqrt{p}\mb{w}\mb{1}-\sqrt{p}\mb{1}^T\mb{w}^H\nonumber\\&+\mb{w}\mb{B}\mb{w}^H+p\mb{w}\mb{S}\mb{w}^H+1,
\end{align}
where $\mb{1}$ represents an all-one $L\times 1$ vector, and the matrices $\mb{B}$ and $\mb{S}$ are respectively
\begin{equation}
    \mb{B}=\left[\begin{array}{cccc}
         \frac{\sigma^2}{|\hat{h}_1|^2}& 0 & ... & 0  \\
         0 & \sigma^2\mathbb{E}|\frac{1}{\hat{h}_2}|^2 & ... & 0\\
         \vdots & \vdots & \ddots & \vdots\\
         0 & 0 & ... & \sigma^2\mathbb{E}|\frac{1}{\hat{h}_L}|^2
    \end{array} \right],
\end{equation}
and 
\begin{align}
    \mb{S}=\diag\Bigg[&\mathbb{E}\left|{\tilde{h}_1}\right|^2\frac{1}{|\hat{h}_1|^2},\mathbb{E}\left|{\tilde{h}_2}\right|^2\mathbb{E}\left|\frac{1}{\hat{h}_2}\right|^2,\nonumber\\&...,\mathbb{E}\left|{\tilde{h}_L}\right|^2\mathbb{E}\left|\frac{1}{\hat{h}_L}\right|^2\Bigg],
\end{align}
Using the first-order optimality condition, the optimal solution for the optimization problem in (\ref{optimization 2}) is given by 
\begin{equation}
    \mb{w}=\sqrt{p}\mb{1}^T\left(p\mb{1}\mb{1}^T+p\mb{S}+\mb{B}\right)^{-1}.
\end{equation}

The capacity in this scenario is given by the mutual information between the transmitted signal $s$ and the final centrally-processed signal $\hat{s}$, as in (\ref{mutual information}), where
$H(\hat{s})\leq \log(\pi e (p\mb{w}\mb{1}\mb{1}^T\mb{w}^H+\mb{w}\mb{B}\mb{w}^H))$ and $H(\hat{s}|s)=\log(\pi e \mb{w}\mb{B}\mb{w}^H)$. Thus, the achievable rate for the partial CSI case can be written as
\begin{equation}\label{rate PC}
R_{\text{PC}}=\log \left( 1+\frac{p\mb{w}\mb{1}\mb{1}^T\mb{w}^H}{\mb{w}\mb{B}\mb{w}^H} \right).
\end{equation}
Note that this capacity is maximized when perfect CSI is available, i.e., $\mb{\hat{h}}=\mb{h}$. For estimated values of CSI, the given rate is reduced.

\section{Numerical Results}\label{Numerical Results}
\begin{figure}
    \centering
    \includegraphics[scale=0.13]{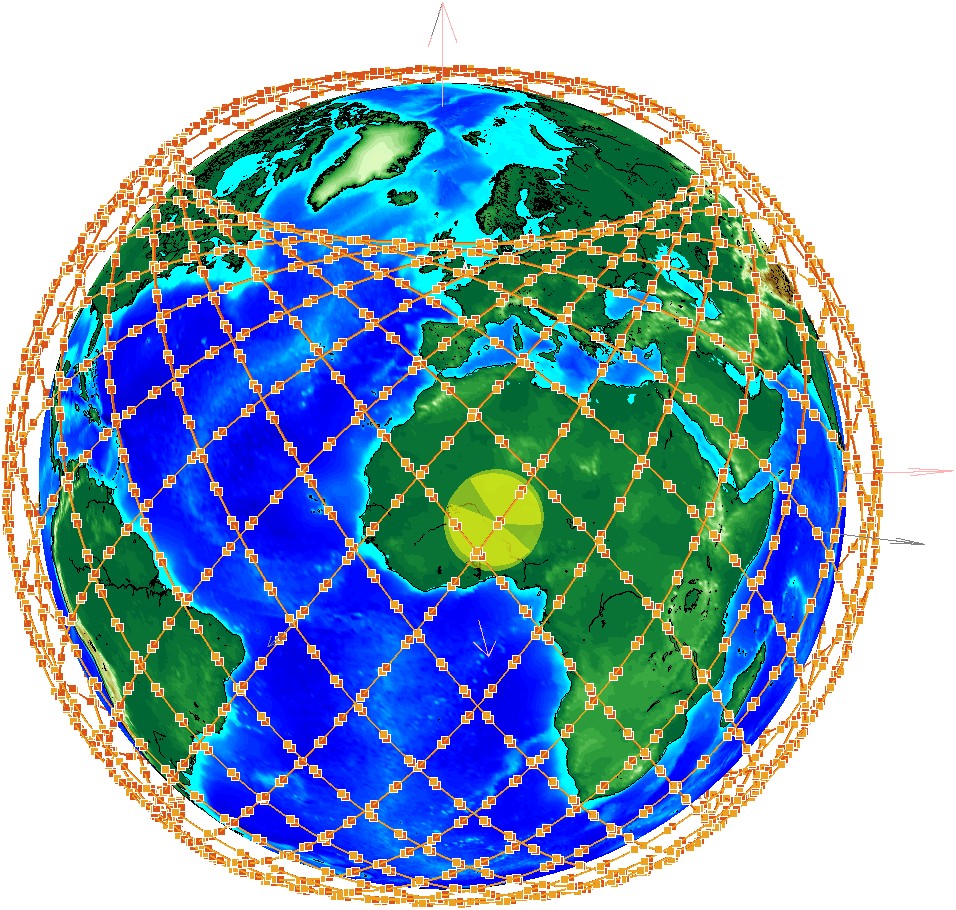}
    \caption{Constellation}
    \label{fig:Constellation}
\end{figure}

In this section, we present the results for both cases in section III. For the simulations we adopt the Starlink constellation with two different layers corresponding to two groups in Starlink launching plans \cite{liang2021phasing}. However, we only use the coordinates of the constellation and not the specifics of the Starlink satellites. Each group has $22$ orbital planes spaced $16.4^{\degree}$ apart with $72$ satellites per plane. 
The first group has $1584$ satellites orbiting at an altitude of $550$ km with a $53$-degree inclination and the second group has $1584$ satellites orbiting at an altitude of $540$ km with an inclination of $53.2^{\degree}$. The true anomaly phasing is $1.1364^{\degree}$, the RAAN spread is $ 360^{\degree}$, and the minimum elevation angle for satellite visibility is $30^{\degree}$. Figure \ref{fig:Constellation} depicts the constellation where there are a total of $3168$ satellites in orbit.
The uplink communication is performed using the C-band, with a carrier frequency of $f_c=6$ GHz and a bandwidth of $BW=500$ MHz. The transmitter device is assumed to be a mobile UH with a transmit power of $P_T=-2$ dBW and a transmit antenna gain of $G_T=5$ dB. The minimum receiver antenna gain is assumed to be $G_R=35$ dB. The simulation is conducted using the approximate user location of London with a latitude of $51.4880572$ degrees and a longitude of $-0.0762838$ degrees. In this constellation, at least $28$ satellites are within the visibility region of the user in the given location. The channel estimation error distribution in (\ref{estimation error distribution}) is assumed to be $\sigma_{\tilde{h}_m}^2=\epsilon^2\text{var}({h}_m)$, with a parameter $\epsilon=3$. Our simulations evaluate the system's performance using uncoded BPSK modulation.

\begin{figure}[t]
    \centering
    \includegraphics[scale=0.19]{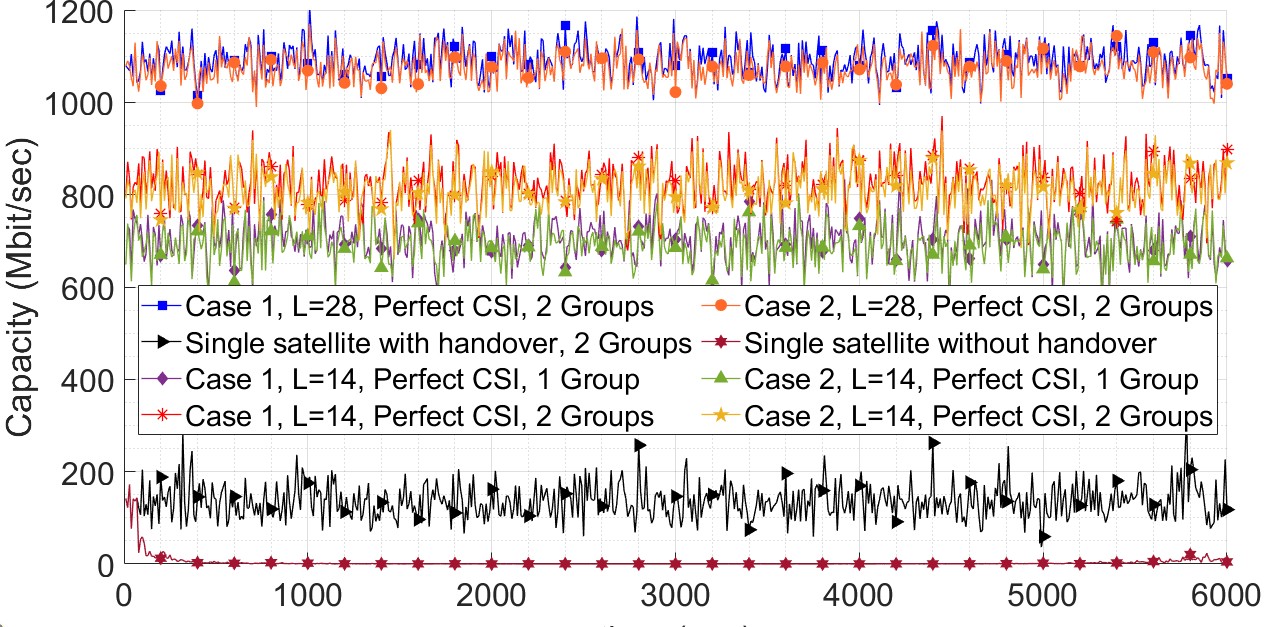}
    \caption{Average capacity in a 100-minute duration of time. }
    \label{fig:1}
\end{figure}

Figure \ref{fig:1} presents the capacity of the system over a duration of $6000$ sec, for both cases. 
The results indicate that employing both groups of satellites in the constellation, with $L=28$ satellites per cluster, leads to a capacity gain exceeding $1000$ Mbit/sec. Furthermore, by choosing only  $L=14$ nearest satellites, a capacity of over $800$ Mbit/sec can be attained. When only one group of satellites in the constellation is used, with $L=L_{max}=14$, the capacity is approximately $700$ Mbit/sec. This capacity reduction can be attributed to the constellation's density. Also, this figure demonstrates that having more satellites cooperating improves system performance at the expense of system overhead.
As expected, Case 2 with partial CSI shared in a cluster exhibits slightly lower capacity than Case 1 with full CSI. The capacity of a single satellite without handover during this time period is also depicted here, which peaks for approximately one minute while the satellite is within the visibility region for nearly five minutes. This suggests that, handovers would be needed every minute in non-cooperative systems to achieve the capacity of the nearest satellite. Thus, satellite cooperation can significantly enhance system performance compared to conventional satellite communication schemes.

\begin{figure}[t]
    \centering
    \includegraphics[scale=0.38]{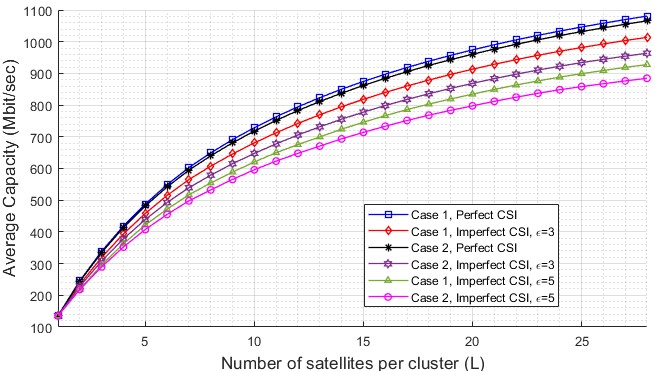}
    \caption{Average capacity vs. number of satellites in a cluster.}
    \label{fig:2}
\end{figure}

Figure \ref{fig:2}, depicts the average capacity of the system as a function of $L$, in both cases, for different channel uncertainty distributions, $\epsilon=3 \ \text{and} \ 5$. As expected, increasing the number of satellites engaged in cooperative detection enhances the system capacity.
However, the capacity does not increase linearly due to the selection of satellites in a cluster based on their proximity to the user. Incorporating distant satellites may not yield significant performance improvements and could lead to high overhead.
For instance, assuming perfect CSI, Case 1 achieves a capacity of $800$ Mbit/sec with $L=12$ satellites. 
However, doubling the number of satellites to $L=24$ results in a capacity increase to $1.04$ Gbit/sec, representing a mere $25\%$ improvement despite the doubled satellite count.
Hence, selecting an optimal number of satellites in a cluster is crucial for maximizing system capacity while minimizing overhead. The optimization of constellation parameters is essential to strike the right balance between capacity and cost, making it an intriguing area for future research.

\begin{figure}[t]
    \centering
    \includegraphics[scale=0.47]{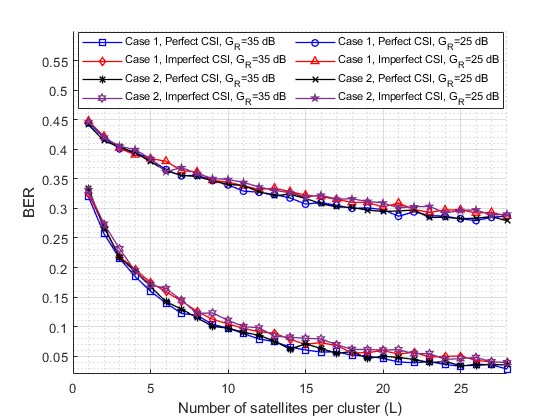}
    \caption{BER  vs. number of satellites in a cluster. }
    \label{fig:3}
\end{figure}

Figure \ref{fig:3}, demonstrates the BER of the system as a function of $L$, in both cases, for different receiver gains, $G_R=35 \ \text{and} \ 20 \ \text{dB}$. Note that the BER results are achieved without using any coding; thus, the actual BER when coding gain is applied would be much lower. As expected, higher values of  $G_R$ can compensate for the path loss and enhance system performance. It is noteworthy to mention that the antenna gains affect the slope of the BER reduction. 

\begin{figure}[t]
    \centering
    \includegraphics[scale=0.31]{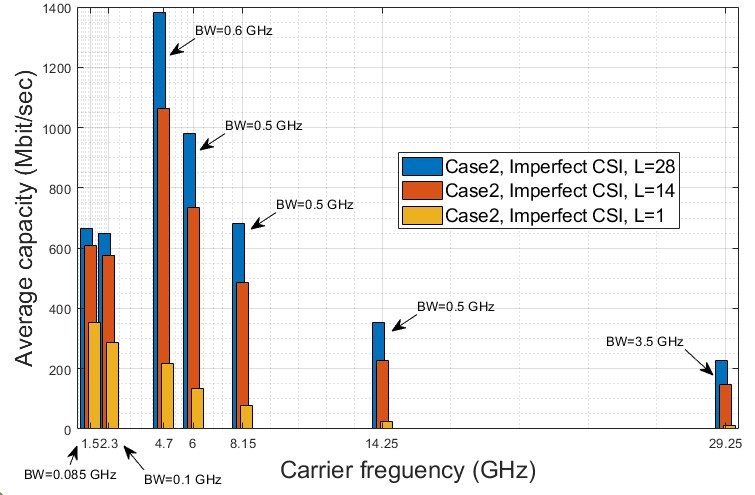}
    \caption{Average Capacity in different $f_c$ and $BW$.  }
    \label{fig:4}
\end{figure}

In Figure \ref{fig:4}, the average capacity of the system is presented in different carrier frequencies and bandwidths. Note that the chosen values for frequency and bandwidth are selected from the 3GPP standard for 5G and {the predefined bands for satellite uplink.\footnote{The frequency band $7.9-8.4$ GHz in X-band is a NATO band type 1 for satellite uplink, the frequency band $14-14.5$ GHz in ku-band is designated for fixed satellite service uplink in North America, and the frequency band $27.5-31$ GHz in ka-band is used for uplink satellite communications. }}
The second case is analyzed in this figure in terms of average capacity under a scenario of imperfect CSI, with a value of $\epsilon=3$ and $G_R=35$ dB.
By using both groups of satellites in the constellation, we compare the system when $L=28$, $L=14$ and $L=1$.
{It is a well-known fact that increasing the bandwidth or decreasing the frequency leads to an increase in capacity. However, it is difficult to predict the behavior of capacity when both parameters are changed simultaneously and thus requires evaluation. 
This figure demonstrates that even in ka-band, utilizing the cooperative satellite technique enables achieving a capacity of approximately $200$ Mbit/sec. In contrast, conventional satellite communication yields a lower capacity of less than $20$ Mbit/sec. 
In lower frequency bands on the other hand, the difference between the capacity of single-satellite scenario where $L=1$ and the cooperative satellite scenarios where $L=14$ and $28$ is much smaller. This is due to the fact that in higher frequencies where path-loss is high, in the formulas for capacity presented in both (\ref{rate FC}) and (\ref{rate PC}), the second term in logarithm becomes much smaller than 1 and thus the capacities can be approximated by $R_{\text{FC}}\approx\frac{p\mb{v}\mb{h}\mb{h}^H\mb{v}^H}{\sigma^2\mb{v}\mb{v}^H}$ and $R_{\text{PC}}\approx\frac{p\mb{w}\mb{1}\mb{1}^T\mb{w}^H}{\mb{w}\mb{B}\mb{w}^H}$, respectively. This means that in higher frequencies the capacity increases with $L$ linearly, while in lower frequencies, the increase in capacity is logarithmic.
Additionally, it is evident that at lower frequencies, the highest capacity is obtained when the bandwidth is wide. It is important to note that these findings may vary depending on the value of the parameter $G_R$.}
\section{conclusion}\label{conclusion}
{This paper explores the uplink transmission of a single-antenna mobile user to a cluster of satellites that collaborate to detect the received signal. Two cases were considered: one with full CSI and the other with partial CSI shared among the satellites. The latter imposed less overhead on the system, while the performance loss was found to be negligible compared to full CSI sharing.
Numerical results demonstrated that our cooperative designs outperformed conventional satellite communication schemes, where only one satellite was able to connect to a user. The study also highlighted that the performance of the system is affected not only by the number of satellites in a cluster but also by constellation parameters such as satellite density. Furthermore, the results underscored the significance of frequency and bandwidth evaluation in system design and emphasized the importance of satellite cooperation in higher frequency bands.}

\section*{Acknowledgment}
\vspace{-0.09cm}
This work is produced in the framework of the Surrey-Huawei joint-lab project. We would also like to acknowledge the support of the University of Surrey 5GIC \& 6GIC.

\appendices
\section{}\label{FirstAppendix}
Here we provide a proof for (\ref{5}). Given  $\mathbb{E}|s|^2=1$, we have
\begin{align}
    \text{MSE}=&\mathbb{E}\{|\hat{s}-s|^2\}=\mathbb{E}\{(\mb{v}\mb{y}-s)(\mb{y}^H\mb{v}^H-s^*) \}\nonumber\\&=
    p\mb{v}\mb{\hat{h}}\mb{\hat{h}}^H\mb{v}^H-2\re\{\sqrt{p}\mb{\hat{h}}^H\mb{v}^H\}\nonumber\\&\quad+p\mb{v}\mathbb{E}\{\mb{\Tilde{h}}\mb{\Tilde{h}}^H\}\mb{v}^H+\mb{v}\mathbb{E}\{\mb{n}\mb{n}^H\}\mb{v}^H+1.
\end{align}
Now we calculate the gradient, w.r.t $\mb{v}^H$ as
\begin{align}
    \frac{\partial \text{MSE}}{\partial \mb{v}^H}=p\mb{v}\mb{\hat{h}}\mb{\hat{h}}^H-\sqrt{p}\mb{\hat{h}}^H+p\mb{v}\mathbb{E}\{\mb{\Tilde{h}}\mb{\Tilde{h}}^H\}+\mb{v}\mathbb{E}\{\mb{n}\mb{n}^H\}.
\end{align}
We have $\Sigma_h=\mathbb{E}\{\mb{\Tilde{h}}\mb{\Tilde{h}}^H\}$ and $\mathbb{E}\{\mb{n}\mb{n}^H\}=\sigma^2\mb{I}$. Thus by setting the gradient equal to zero, the proof is completed.

Now, similarly, a proof for (\ref{rate FC}) is provided.
We have
\begin{align}
    H(\hat{s})\leq \log(\pi e R_{\hat{s}}),
\end{align}
where $R_{\hat{s}}=\mathbb{E}\{\hat{s}\hat{s}^*\}$. Similar to proof of (\ref{5}), we have
\begin{align}
    R_{\hat{s}}&=\mathbb{E}\{\mb{v}\mb{y}\mb{y}^H\mb{v}^H\}
    =p\mb{v}\mb{\hat{h}}\mb{\hat{h}}^H\mb{v}^H+\mb{v}\mathbb{E}\{\mb{n}\mb{n}^H\}\mb{v}^H\nonumber\\&=p\mb{v}\mb{\hat{h}}\mb{\hat{h}}^H\mb{v}^H+\sigma^2\mb{v}\mb{v}^H.
\end{align}
The expression $H(\hat{s}|s)=\log(\pi e \sigma^2\mb{v}\mb{v}^H)$ can be proved in the same manner. Thus, the proof is completed.

\bibliographystyle{IEEEtran}
\bibliography{SatelliteRef}
\end{document}